\documentclass[conference,letterpaper]{IEEEtran}
\IEEEoverridecommandlockouts

\usepackage[utf8]{inputenc} 
\usepackage[T1]{fontenc}
\usepackage{url}
\usepackage{ifthen}
\usepackage{cite}
\usepackage[cmex10]{amsmath}
\usepackage{tikz}
\usetikzlibrary{quantikz2}
\usepackage{comment}
\usepackage{cleveref}
\usepackage{amsmath}
\usepackage{amssymb}
\usepackage{amsfonts}
\usepackage{thmtools}
\declaretheorem{definition}
\declaretheorem[sibling=definition]{theorem}

\declaretheorem[sibling=definition]{proposition}
\declaretheorem[sibling=definition]{lemma}
\declaretheorem[sibling=definition]{remark}
\DeclareMathOperator{\tr}{Tr}
\DeclareMathOperator{\conv}{conv}

\DeclareMathOperator{\poly}{poly}

\DeclareMathOperator{\LDP}{LDP}
\allowdisplaybreaks

\begin{document}
\title{Sample Complexity of Composite Quantum Hypothesis Testing\\
\thanks{STJ has received funding from EPSRC Quantum Technologies Career Acceleration Fellowship (UKRI1218).}}


\author{%
  \IEEEauthorblockN{Jacob Paul Simpson, Efstratios Palias and Sharu Theresa Jose}
  \IEEEauthorblockA{School of Computer Science,
                    University of Birmingham, UK\\
                    Email: jps538@student.bham.ac.uk,
                    e.palias@bham.ac.uk,
                    s.t.jose@bham.ac.uk}
}
\maketitle
\begin{abstract}
This paper investigates symmetric composite binary quantum hypothesis testing (QHT), where the goal is to determine which of two uncertainty sets contains an unknown quantum state. While asymptotic error exponents for this problem are well-studied, the finite-sample regime remains poorly understood. We bridge this gap by characterizing the sample complexity -- the minimum number of state copies required to achieve a target error level. Specifically, we derive lower bounds that generalize the sample complexity of simple QHT and introduce new upper bounds for various uncertainty sets, including of both finite and infinite cardinalities. Notably, our upper and lower bounds match up to universal constants, providing a tight characterization of the sample complexity. Finally, we extend our analysis to the differentially private setting, establishing the sample complexity for privacy-preserving composite QHT.
\end{abstract}

\section{Introduction}
Hypothesis testing is a fundamental problem in statistical inference where the goal is to determine which category a test object belongs to. Quantum hypothesis testing (QHT) extends this paradigm to the quantum realm, merging quantum mechanics with mathematical statistics, to address fundamental tasks in quantum information processing and communication \cite{barndorff2003quantum,bae2015quantum}. In its simplest form, a simple binary QHT problem is a state discrimination problem: given $n$ copies of an unknown quantum state, a distinguisher must design a two-outcome quantum measurement to identify which of the two known states, $\rho$ or $\sigma$, was prepared \cite{Helstrom1969,Holevo1973}. The performance of such a test is typically evaluated through two lenses: symmetric QHT, which seeks to minimize the average probability of error, and asymmetric QHT, which  minimizes the type-II error subject to a fixed constraint on the type-I error \cite{cheng2025invitation}. 

While simple QHT involves distinguishing between individual states, many practical scenarios involve \emph{composite binary QHT}. In this setting, the goal is to determine which of two \emph{uncertainty sets} of quantum states contains the unknown state. This generalizes the problem to more complex, real-world conditions, where the exact state may not be known precisely, due to experimental limitations or environmental noise. 


\enlargethispage{-.05in}Existing research in both simple and composite QHT has primarily focused on asymptotic error exponents, which characterize the exponential rate of error decay as the number of state copies $n$ approaches infinity. For instance, the quantum Chernoff exponent for symmetric simple binary QHT was established in \cite{audenaert2008asymptotic, nussbaum2009chernoff} and recently generalized to the composite setting in \cite{fang2025generalized}. Similarly, for asymmetric QHT, the quantum Stein exponent has been studied for simple QHT in \cite{hiai1991proper ,ogawa2002strong} and has been generalized to the composite setting under structural assumptions on the uncertainty sets in \cite{hayashi2002optimal,  brandao2010generalization, berta2021composite, fang2024generalized, hayashi2025generalized}. These results provide fundamental limits in the regime of infinite resources but do not fully capture the requirements of practical, resource-constrained quantum systems. 

Recently, an alternative line of research has emerged to study QHT in the non-asymptotic setting by  analysing the \emph{sample complexity}. This refers to the minimum number of quantum state copies, $n^\ast(\delta)$, required to achieve a desired target error level $\delta$. For simple symmetric QHT between quantum states $\rho_1$ and $\rho_2$, recent work \cite{cheng2025invitation} has characterized this complexity as $n^\ast(\delta)= \Theta\Bigl( \frac{\ln(1/\delta)}{-\ln F(\rho_1,\rho_2)}\Bigr)$ where $F(\rho_1,\rho_2)$, denotes the (square root) Uhlmann fidelity between the quantum states. This analysis has since been extended to simple QHT under local differential privacy constraints \cite{Cheng2024LDPQHT, nuradha2025contraction}, where the unknown quantum state is affected by a differentially private quantum channel before reaching the tester.

\enlargethispage{-.05in}Building on these recent developments, this paper provides the first comprehensive characterization of the \textbf{sample complexity of symmetric composite QHT}. Our key contributions are as follows: First, we derive new lower and upper bounds on the sample complexity for various classes of uncertainty sets, including both finite and infinite cardinalities. Specifically, for testing a singleton pure uncertainty set $\mathcal{D}_1=\{\proj\psi\}$ against a composite uncertainty set $\mathcal{D}_2$, we show that the sample complexity scales as $n^\ast(\delta)=\Theta \Bigl(\frac{\ln(1/\delta)}{-\ln \sup_{\rho_2 \in \mathcal{D}_2} \bra\psi \rho_2 \ket\psi} \Bigr)$. We generalize this result to the case when both the sets are finite and show that the sample complexity is governed by the maximum pairwise fidelity between the sets, $F_{\max}:=\sup_{\rho_i \in \mathcal{D}_i, i=1,2}F(\rho_1,\rho_2)$, as 
    $\Omega\Bigl( \frac{\ln(1/\delta)}{-\ln F_{\max}}\Bigr) \leq n^\ast(\delta)\leq \mathcal{O}\Bigl(\frac{\ln(\sqrt{\vert\mathcal{D}_1\vert\,\vert\mathcal{D}_2\vert}/\delta)}{-\ln F_{\max}}\Bigr)$.
    For infinite cardinality uncertainty sets, we show that under the constraint $F_{\max}\leq c$, for some $c \in (0,1)$,  the sample complexity scales as $ n^\ast(\delta)= \Theta \Bigl(\frac{\ln(1/\delta)}{-\ln F_{\max}}\Bigr)$. Secondly, we extend our framework to the locally differentially private setting, establishing the first sample complexity bounds for private composite QHT.

\section{Problem Setting}\label{sec:problem}
\subsection{Notation} 
For any bounded operator $A$, we denote $\tr(A)$ as the trace of $A$, with its trace norm defined as $\|A\|_1:=\tr(\sqrt{A^\dagger A})$. Throughout this paper, we use (indexed) $\rho$ and $\sigma$ to denote quantum states represented as density matrices, i.e., positive semi-definite, unit-trace, Hermitian matrices, acting on a Hilbert space of dimension $d$. Then, $F(\rho,\sigma):=\|\sqrt \rho\sqrt \sigma\|_1$ defines the (square root) Uhlmann fidelity, and $d_{B}(\rho,\sigma):= \sqrt{\,2\left(1 - F(\rho,\sigma)\right)}$ defines the Bures distance between the two states. We use $\dim_{\mathbb R}(\cdot)$ to denote the \emph{real} dimension of the smallest affine space generated by the set `$\cdot$'. If the set is a vector space, then $\dim_{\mathbb R}(\cdot)$ coincides with its standard linear dimension, i.e., the number of linearly independent vectors whose \emph{real} span is the vector space.

\subsection{Simple Quantum Hypothesis Testing}
In simple binary QHT, a quantum system is prepared in one of two known states $\rho_1^{\otimes n}$ or $\rho_2^{\otimes n}$. The distinguisher is given access to $n$ identical copies of the state without \emph{a priori} knowledge of which was prepared.  
The objective of the distinguisher is to determine the identity of the state by designing a two-outcome  positive operator-valued measurement (POVM) $\{\Pi, I- \Pi\}$, where the outcome associated with the operator $0 \leq \Pi \leq I$ corresponds to guessing $\rho_1^{\otimes n}$ and  the outcome associated with $I-\Pi$ corresponds to guessing $\rho_2^{\otimes n}$.

The probability of error is determined by the Born rule. Specifically, the probability of erroneously guessing $\rho_1$ when the prepared state is  $\rho_2$ is  $\tr(\Pi \rho_2^{\otimes n})$, while the converse error is  $\tr((I-\Pi) \rho_1^{\otimes n})$. 
Assuming  prior probabilities $p$ and $(1-p)$, where $p \in (0,1)$, for preparing the unknown state in state $\rho_1$ and $\rho_2$ respectively, the \emph{expected  error probability} of simple, symmetric QHT for a given POVM $\{\Pi,I-\Pi\}$ is defined as
\begin{equation}
    P_e(\Pi,p,\rho_1^{\otimes n},\rho_2^{\otimes n}):=p\tr((I-\Pi)\rho_1^{\otimes n})+(1-p)\tr(\Pi\rho_2^{\otimes n}). \label{eq:simple_error_probability}
\end{equation}
The optimal performance is found by minimizing over all possible POVMs. This minimum expected error probability is achieved by the Holevo-Helstrom measurement \cite{Helstrom1969,Holevo1973} as
\begin{align}
P_{e,\min}(p,\rho_1^{\otimes n},&\rho_2^{\otimes n}):=\inf_{0\preceq\Pi\preceq I}P_e(\Pi,p,\rho_1^{\otimes n},\rho_2^{\otimes n})\\
    &= \tfrac12\left( 1 - \left\| p\rho_1^{\otimes n} - (1-p)\rho_2^{\otimes n} \right\|_1 \right). \label{eq:prob_error_simple}
\end{align}

We now define the sample complexity for simple QHT.
\begin{definition}[Sample Complexity of Simple QHT]
The $\delta$-sample complexity $n^\ast(p,\rho_1,\rho_2,\delta)$ for distinguishing between states $\rho_1$ and $\rho_2$ refers to the smallest number of quantum state copies required to achieve a minimum expected error probability of at most $\delta \in (0,1)$, i.e.,
\begin{equation*}\label{eq:sample_complexity}
    n^\ast(p,\rho_1,\rho_2, \delta):=\inf\{n\in\mathbb N:P_{e,\min}(p,\rho_1^{\otimes n},\rho_2^{\otimes n})\leq\delta\}.
\end{equation*}
\end{definition}

For singleton quantum states $\rho_1$ and $\rho_2$, recent work \cite{cheng2025invitation} establishes the following bounds on the sample complexity.
\begin{theorem}[\cite{cheng2025invitation}]\label{th:singleton}
    The following bounds hold for the sample complexity of simple QHT:
    \begin{align}
        \max \biggl \lbrace \frac{\ln\bigl(\frac{p(1-p)}\delta\bigr)}{-2\ln F(\rho_1,\rho_2) }, &\frac{1-\frac{\delta(1-\delta)}{p(1-p)}}{d_B(\rho_1,\rho_2)^2}\biggr \rbrace \leq n^\ast(p,\rho_1,\rho_2,\delta)\nonumber \\ &\leq \biggl \lceil \inf_{s \in [0,1]} \frac{\ln \bigl( \frac{p^s(1-p)^{1-s}}{\delta}\bigr)}{-\ln\tr(\rho_1^s \rho_2^{1-s})} \biggr \rceil. \label{eq:samplecomplexity_simpleQHT}
    \end{align}
\end{theorem}
\enlargethispage{-.05in}We next generalize this setting to composite hypotheses sets.
\subsection{Composite Quantum Hypothesis Testing}
In composite QHT, the distinguisher tests if the unknown quantum state belongs to either of the two known \emph{uncertainty} sets of quantum states: $\mathcal{D}_{1,n}$ or $\mathcal{D}_{2,n}$.  We define these sets as
\begin{equation}
    \mathcal D_{i,n}:=\{\rho_i^{\otimes n}:\rho_i\in\mathcal D_i\}, \quad \mbox{for} \hspace{0.1cm} i=1,2, \label{eq:uncertaintysets}
\end{equation} consisting of $n$-independent copies of quantum states drawn from compact sets
 $\mathcal D_1$ or $\mathcal D_2$ of density operators in a $d$-dimensional Hilbert space \cite{berta2021composite}. While simple QHT distinguishes between two fixed states, the composite framework requires discriminating between two sets of quantum states; it reduces to the simple case only  when $\mathcal{D}_i$ are of unit cardinality.

\enlargethispage{-.05in}In composite QHT, the distinguisher aims to design a two-outcome POVM $\{\Pi, I-\Pi\}$ that minimizes the probability of error, where the operator $\Pi$ corresponds to the hypothesis that the unknown state belongs to $\mathcal{D}_{1,n}$ and  $I-\Pi$ to $\mathcal{D}_{2,n}$. 
 However, unlike the simple QHT, the exact state within the uncertainty sets in which the quantum system is prepared is unknown, necessitating a  worst-case approach. Therefore, we define the composite expected error probability of a given POVM as the supremum of the error probability over all possible state pairs within the uncertainty sets: \cite{fang2025generalized} 
 \begin{equation}
    P_e(\Pi,p,\mathcal D_{1,n},\mathcal D_{2,n}):=\sup_{\rho_i\in\mathcal D_i, i=1,2}P_e(\Pi,p,\rho_1^{\otimes n},\rho_2^{\otimes n}), 
\end{equation}where $P_e(\Pi,p,\rho_1^{\otimes n},\rho_2^{\otimes n})$ is defined as in \eqref{eq:simple_error_probability} and $(p,1-p)$ are the prior probabilities for each uncertainty set. The \emph{min-max  expected error probability} is then obtained as
\begin{equation}
    P_{e,\min}(p,\mathcal D_{1,n},\mathcal D_{2,n}):=\inf_{0\preceq\Pi\preceq I}P_e(\Pi,p,\mathcal D_{1,n},\mathcal D_{2,n}). \label{eq:prob_error_composite}
\end{equation}

We can now extend the definition of sample complexity to composite QHT as follows.
\begin{definition}[Sample Complexity of Composite QHT]
The $\delta$-sample complexity $n^\ast(p,\mathcal D_{1},\mathcal D_{2},\delta)$ for distinguishing between sets $\mathcal{D}_1$ and $\mathcal{D}_2$ is defined as
\begin{equation*}
    n^\ast(p,\mathcal D_{1},\mathcal D_{2},\delta):=\inf\{n\in\mathbb N:P_{e,\min}(p,\mathcal D_{1,n},\mathcal D_{2,n})\leq\delta\}.
\end{equation*}
\end{definition}

 In the following sections, we provide the first study of the sample complexity of composite QHT. This analysis unveils how the sample complexity depends on key problem parameters, including the ``similarity" of the uncertainty sets, the number of elements they contain, and the dimensionality of the underlying Hilbert space. 
 For notational convenience, we abbreviate
$n^\ast(p,\mathcal D_{1},\mathcal D_{2},\delta)$ (or $n^\ast(p,\rho_{1},\rho_{2},\delta)$) as $n^\ast(\delta)$ whenever the prior and the sets (or states) are clear from the context.
Finally, we exclude the following trivial cases from our analysis, which generalizes the regimes identified in \cite[Remark 2]{cheng2025invitation}. A proof can be found in Appendix \ref{Proof of  trivial cases}.
\begin{remark}\label{trivial_cases}
     If $\mathcal D_1\perp\mathcal D_2$, or $\delta \in [\tfrac{1}{2},1]$, or there exists an $s\in[0,1]$ such that $\delta\geq p^s(1-p)^{1-s}$, then $n^\ast(\delta)=1$. Also, if $\mathcal D_1\cap\mathcal D_2\neq\emptyset$ and $\delta<\min\{p,1-p\}$, then $n^\ast(\delta)=+\infty$.
\end{remark}

\section{Main Results: Sample Complexity Bounds}\label{sec:main}
In this section, we present our results, establishing lower and upper bounds on the sample complexity of composite QHT. We start by analysing the error probability, and use the insights as building blocks to derive the main bounds.
\subsection{Analysis of Min-max Expected Error Probability}
We first provide an alternative characterization of the min-max expected error probability \eqref{eq:prob_error_composite}. To this end, we define the convex hulls of the uncertainty sets $\mathcal{D}_{i,n}$ in \eqref{eq:uncertaintysets} as
\begin{equation}
    \mathcal C_{i,n}:=\conv(\mathcal D_{i,n})=\left\{\int\rho_i^{\otimes n}d\mu_i(\rho_i):\mu_i\in\mathcal P(\mathcal D_i)\right\}, \label{eq:convexhull}
\end{equation}
where $\mathcal P(\mathcal D_i)$ is the set of all probability measures on $\mathcal D_i$, for $i=1,2$. We then have the following result.
\begin{proposition}\label{prob_error_composite_explicit}
   The following relationship holds:
 \begin{multline}
     P_{e,\min}(p,\mathcal D_{1,n},\mathcal D_{2,n})=\\\sup_{\sigma_{i,n}\in\mathcal C_{i,n},i=1,2}\left(\tfrac12-\tfrac12\|p\sigma_{1,n}-(1-p)\sigma_{2,n}\|_1\right),\label{eq:prob_error_composite_explicit}
 \end{multline}
 where $\mathcal{C}_{i,n}$ is defined as in \eqref{eq:convexhull}.
\end{proposition}
\begin{IEEEproof}
   Starting from the definition in \eqref{eq:prob_error_composite}, the following series of relationships hold:
\begin{align*}
P_{e,\min}(p,\mathcal D_{1,n},\mathcal D_{2,n})
        &=\inf_{0\preceq\Pi\preceq I}\sup_{\sigma_{i,n} \in \mathcal{C}_{i,n}}P_e(\Pi,p,\sigma_{1,n},\sigma_{2,n})\\
        &=\sup_{\sigma_{i,n} \in \mathcal{C}_{i,n}}\inf_{0\preceq\Pi\preceq I}P_e(\Pi,p,\sigma_{1,n},\sigma_{2,n})\\
        &=\sup_{\sigma_{i,n} \in \mathcal{C}_{i,n}}P_{e,\min}(p,\sigma_{1,n},\sigma_{2,n}),
\end{align*}
where in the first equality, we used linearity of $P_e$ in its argument quantum states, whereby the supremum is achieved at the extreme points of the convex hull. The second equality follows from applying Sion's minimax theorem \cite[Lemma A.1]{berta2021composite} to swap the order of $\sup$ and $\inf$, since $P_e(\Pi,p,\sigma_{1,n},\sigma_{2,n})$ is linear in $\Pi$ and  $\mathcal{C}_{i,n}$ are compact and convex sets. The result then follows from the last equality via \eqref{eq:prob_error_simple}.
\end{IEEEproof}


Importantly, the following lemma shows that the error probability in \Cref{prob_error_composite_explicit} decreases in the number of copies.  \begin{lemma}\label{lemma:error probability is non-increasing in n}
        Let $\Delta_n:=\inf_{\sigma_{i,n}\in\mathcal C_{i,n}}\|p\sigma_{1,n}-(1-p)\sigma_{2,n}\|_1$. Then, for all $n\in\mathbb N$, we have
        $
            \Delta_n\leq\Delta_{n+1}$ and consequently, $P_{e,\min}(p,\mathcal D_{1,n},\mathcal D_{2,n}) \geq P_{e,\min}(p,\mathcal D_{1,n+1},\mathcal D_{2,n+1})$. 
\end{lemma}
\begin{IEEEproof}
The proof follows by using the partial trace over the $(n+1)$-th copy, $\tr_{n+1} (\cdot)$, which is a completely positive trace-preserving (CPTP) map, and noting that $\tr_{n+1}(\mathcal C_{i,n+1})=\mathcal C_{i,n}.$ Then, for minimizers $\sigma_{i,n+1}^*\in\mathcal C_{i,n+1}$, $i=1,2$, of $\Delta_{n+1}$, we get
        \begin{align*}
            \Delta_n&\leq \|\tr_{n+1}(p\sigma_{1,n+1}^*-(1-p)\sigma_{2,n+1}^*)\|_1\\
            &\leq\|p\sigma_{1,n+1}^*-(1-p)\sigma_{2,n+1}^*\|_1\\
            &=\inf_{\sigma_{i,n+1}\in\mathcal C_{i,n+1}}\|p\sigma_{1,n+1}-(1-p)\sigma_{2,n+1}\|_1=\Delta_{n+1},
        \end{align*}
       where the first inequality follows because $\tr_{n+1}(\sigma_{i,n+1}^*)\in\mathcal C_{i,n}$ and the second inequality uses that $\|\tr_{n+1}(A)\|_1\leq\|A\|_1$ for a CPTP map acting on a Hermitian operator $A$.
\end{IEEEproof}

We next present a series of upper bounds on the min-max expected error probability in  \eqref{eq:prob_error_composite_explicit}, which will be used later.
\begin{proposition}
    The following series of upper bounds hold:
    \begin{align}
      &P_{e,\min}(p,\mathcal D_{1,n},\mathcal D_{2,n})  \nonumber \\&\leq\sup_{\sigma_{i,n}\in\mathcal C_{i,n}, i=1,2}\inf_{0\leq s\leq1}p^s(1-p)^{1-s}\tr(\sigma_{1,n}^s\sigma_{2,n}^{1-s})\label{eq:quantum_chernoff_bound}
      \end{align} \begin{align}
    &\leq\sqrt{p(1-p)}\sup_{\sigma_{i,n}\in\mathcal C_{i,n}, i=1,2}\tr(\sqrt{\sigma_{1,n}}\sqrt{\sigma_{2,n}})\\
    &\leq\sqrt{p(1-p)}\sup_{\sigma_{i,n}\in\mathcal C_{i,n}, i=1,2}F(\sigma_{1,n},\sigma_{2,n})\label{eq:fidelity upper bound}.
\end{align}
\end{proposition}
\begin{IEEEproof}
    The first inequality uses the well-known quantum Chernoff Bound \cite[Theorem 1]{Audenaert2007QuantumChernoff}, while the second inequality follows by choosing  $s=1/2$. The last inequality follows from \cite[Eq. 28]{audenaert2008asymptotic} and the definition of Uhlmann fidelity.
\end{IEEEproof}

\subsection{Lower Bounds on Sample Complexity}
We now present lower bounds on the $\delta$-sample complexity of composite QHT, generalizing the results in \cite{cheng2025invitation}. 
We define
\begin{equation}
F_{\max}:=\sup_{\rho_i\in\mathcal D_i, i=1,2}F(\rho_1,\rho_2)
\end{equation} as the maximum fidelity between pairs of quantum states within the uncertainty sets. We then have the following.

\begin{theorem}\label{th:lb_on_sample_complexity}
Let $\mathcal D_1,\mathcal D_2$ be arbitrary compact sets of quantum states. Then, for $p, \delta \in (0,1)$,
\begin{equation}\label{eq:symmetric_lb}
 \max\left\{
   \frac{\ln\bigl(\frac{p(1-p)}{\delta}\bigr)}
    {-2\ln F_{\max}},\\ \frac{1-\frac{\delta(1-\delta)}{p(1-p)}}{\inf_{\rho_i\in\mathcal D_i}d_B^2(\rho_1,\rho_2)}  
\right\}\leq n^\ast(\delta).  
    \end{equation}
\end{theorem}
\begin{IEEEproof}
Since the min-max error probability is defined in a worst-case sense over the uncertainty sets, we have $ n^\ast(\delta,p,\mathcal D_{1}, \mathcal D_{2}) \ge n^\ast(\delta,p,\rho_1,\rho_2)$ for any pair $(\rho_1,\rho_2)$, and hence $ n^\ast(\delta,p,\mathcal D_{1}, \mathcal D_{2}) \ge \sup_{\rho_i\in\mathcal D_i} n^\ast(\delta,p,\rho_1,\rho_2)$. Therefore, the desired result follows from the lower bound in \Cref{th:singleton} by taking the supremum over all pairs of states $(\rho_1,\rho_2)$.
\end{IEEEproof}
\Cref{th:lb_on_sample_complexity} shows that the lower bound of the sample complexity depends on the least distinguishable pair of states from $\mathcal{D}_1 \times \mathcal{D}_2$. For singleton sets, \Cref{th:lb_on_sample_complexity} collapses to the lower bound in \Cref{th:singleton}.
\subsection{Upper Bounds on Sample Complexity}
We now present upper bounds on the sample complexity by considering uncertainty sets of increasing ``complexity''. 
\subsubsection{Quantum State Verification Problem}
Consider the task of quantum state verification (QSV) \cite{Thinh2020worstcase}, where  $\mathcal{D}_1$ consists of a singleton pure state $\proj\psi$, and $\mathcal{D}_2$ is any compact set of quantum states satisfying $\proj\psi\notin\mathcal{D}_2$. The following theorem provides an upper bound on the sample complexity of QSV.
\begin{theorem}\label{prop:sample_complexity_qsvp}
Let $\mathcal D_{1}=\{\proj\psi\}$ be a singleton pure state and $\mathcal D_2$ be a compact set of quantum states satisfying $\mathcal{D}_1 \cap \mathcal{D}_2 = \emptyset$. Then, for $p,\delta \in (0,1)$,
\begin{equation}\label{eq:sample_complexity_qsvp}
   n^{\ast}(\delta) \leq\left\lceil  \frac{ \ln \bigl( \frac{1-p}{\delta}\bigr)}{-\ln \sup_{\rho_2 \in \mathcal{D}_2} \bra{\psi}\rho_2\ket\psi} \right\rceil.
\end{equation}
\end{theorem}
\begin{IEEEproof}
    The proof of \eqref{eq:sample_complexity_qsvp} follows from using the upper bound in \eqref{eq:quantum_chernoff_bound} by noting that $\sigma_{1,n}^s=(\proj\psi^{\otimes n})^s=\proj\psi^{\otimes n}$ for $s \in [0,1]$. Choosing $s=0$,
    then yields that
    \begin{align*}
      P_{e,\min}(p,\mathcal D_{1,n},\mathcal D_{2,n})&\leq
      (1-p)\sup_{\sigma_{2,n} \in \mathcal{C}_{2,n}} \tr(\sigma_{2,n} \proj\psi^{\otimes n})\\&= (1-p)\sup_{\rho_2 \in \mathcal{D}_2} \tr(\rho_{2} \proj\psi)^{n},
    \end{align*}
    where the equality follows since the supremum is attained at an extreme point of the convex hull. The choice of $n$ in \eqref{eq:sample_complexity_qsvp} satisfies $P_{e,\min}(p,\mathcal D_{1,n},\mathcal D_{2,n})\leq \delta$, completing the proof.
\end{IEEEproof}

 This theorem shows that the sample complexity for the QSV problem scales with the maximum overlap between the pure state $\proj\psi$ and the states in the uncertainty set $\mathcal{D}_2$. For fixed $p$, \eqref{eq:symmetric_lb} and \eqref{eq:sample_complexity_qsvp} together imply that the sample complexity for QSV scales as $n^\ast(\delta)=\Theta\Bigl(\frac{\ln(1/\delta)}{-\ln \sup_{\rho_2 \in \mathcal{D}_2} \bra{\psi}\rho_2\ket\psi}\Bigr)$. When $\mathcal{D}_2$ is also singleton,
 the bound \eqref{eq:sample_complexity_qsvp} recovers the sample complexity upper bound of simple QHT in \eqref{eq:samplecomplexity_simpleQHT} with $s=0$ and $\rho=\proj\psi$. We note that the choice of $s=0$ is in fact the minimizer of \eqref{eq:quantum_chernoff_bound} when $p\geq\frac{1}{2}$ (see Appendix \ref{sec:Proof of optimality in prop:sample_complexity_qsvp.} for a proof).
Lastly, we note that the error exponent of QSV problem  evaluates to
 \begin{equation*}
   \hspace{-0.3cm}\lim_{n\rightarrow\infty}\frac{-\ln P_{e,\min}(p,\{\proj\psi\},\mathcal D_{2,n})}{n} = -\!\ln \!\sup_{\rho_2\in \mathcal{D}_2} \bra{\psi}\rho_2\ket\psi, 
 \end{equation*}
recovering the composite quantum Chernoff exponent for $s=0$ \cite{fang2025generalized}. This is consistent  with the known singleton pure–mixed case, where the optimal exponent is attained at $s=0$ \cite{Audenaert2007QuantumChernoff}.

\subsubsection{Uncertainty Sets of Finite Cardinality} We now extend singleton uncertainty sets to uncertainty sets of finite cardinality, i.e.,  $|\mathcal{D}_i|=m_i  < \infty$ for $i=1,2$. To upper bound the sample complexity, we first derive the following result.
\begin{lemma}\label{lemma:fidelity_for_finite}
Let $|\mathcal D_i|=m_i<\infty$ for $i=1,2$. Then,
\begin{equation}\label{ineq:end_of_fidility_chain_finite}
     \sup_{\sigma_{i,n}\in\mathcal C_{i,n}, i=1,2}
    F(\sigma_{1,n},\sigma_{2,n}) \leq \sqrt{m_1m_2}F_{\max}^n.
    \end{equation}
\end{lemma}
\begin{IEEEproof}
  For $i=1,2$, let $\sigma_{i,n}\in\mathcal C_{i,n}$ denote any quantum states in the respective convex hulls, which can be written equivalently as
    $
    \sigma_{1,n} = \sum_{j=1}^{m_1} p_j \rho_{1,j}^{\otimes n} $ and 
    $\sigma_{2,n} = \sum_{k=1}^{m_2} q_k \rho_{2,k}^{\otimes n},
   $
  where $\rho_{1,j}\in\mathcal D_1$, $\rho_{2,k}\in\mathcal D_2$ and  $\{p_j\}, \{q_k\}$ denote probability vectors. For any $\sigma_{i,n} \in \mathcal C_{i,n}$, the following set of relations hold,
    \begin{align*}
F(\sigma_{1,n},\sigma_{2,n})&=F\Biggl(\sum_{j=1}^{m_1} p_j \rho_{1,j}^{\otimes n},\sum_{k=1}^{m_2} q_k \rho_{2,k}^{\otimes n}\Biggr)\\
        &\leq\sum_{j=1}^{m_1}\sum_{k=1}^{m_2}F(p_j\rho_{1,j}^{\otimes n},q_k\rho_{2,k}^{\otimes n})\\
        &=\sum_{j=1}^{m_1}\sum_{k=1}^{m_2}\sqrt{p_jq_k}F(\rho_{1,j},\rho_{2,k})^n\\
        &\leq\sum_{j=1}^{m_1}\sum_{k=1}^{m_2}\sqrt{p_jq_k}\sup_{\rho_i\in\mathcal D_i}F(\rho_1,\rho_2)^n\\
        &\leq\sqrt{m_1m_2}F_{\max}^n \nonumber,
    \end{align*}
where the first inequality follows by sub-additivity of fidelity \cite[Lemma 4.9]{audenaert2014upper}, and the last inequality follows by Cauchy-Schwarz:
 $\sum_{j=1}^{m_1} (\sqrt{p_j}\cdot1)\leq\sqrt{(\sum_{j=1}^{m_1}p_j)m_1}=\sqrt{m_1}.$
\end{IEEEproof}
\enlargethispage{-.1in}We now derive the following upper bound using \Cref{lemma:fidelity_for_finite}.
\begin{theorem}\label{prop:ub_on_sample_complexity_finite}
Let $|\mathcal D_i|=m_i<\infty$ for $i=1,2$. Then, for every $p \in (0,1)$ and $\delta \in (0,1)$,
\begin{equation}
   n^\ast(\delta) \leq \biggl\lceil \frac{\ln \Bigl( \frac{\sqrt{m_1m_2p(1-p)}}{\delta}\Bigr)}{-\ln F_{\max}}\biggr\rceil. \label{eq:ub_finitecardinality}
\end{equation}
Furthermore, for fixed $p$, \eqref{eq:symmetric_lb} and \eqref{eq:ub_finitecardinality} together imply that  $\Omega\Bigl( \frac{\ln(1/\delta)}{-\ln F_{\max}}\Bigr) \leq n^\ast(\delta)\leq \mathcal{O}\Bigl(\frac{\ln(\sqrt{m_1m_2}/\delta)}{-\ln F_{\max}}\Bigr).$ 
\end{theorem}
\begin{IEEEproof}
   The proof of \eqref{eq:ub_finitecardinality} follows by combining the upper bound in \eqref{eq:fidelity upper bound} with \Cref{lemma:fidelity_for_finite} 
   and by verifying that the chosen value of $n$ satisfies $P_{e,\min}(p,\mathcal D_{1,n},\mathcal D_{2,n})\leq \delta$.
\end{IEEEproof}
 
\subsubsection{Uncertainty Sets of Infinite Cardinality} Lastly, we consider the most general setting where the uncertainty sets $\mathcal{D}_i$ are of infinite cardinality and derive upper bounds on the sample complexity. We first extend \Cref{lemma:fidelity_for_finite} to this setting. 
\begin{lemma}\label{lemma:fidelity_for_infinite}
Let $\mathcal D_{1}, \mathcal{D}_2$ be compact sets of quantum states possibly of infinite cardinality. Then,
\begin{equation}\label{eq:fidelity_for_infinite}
\hspace{-0.22cm}\sup_{\sigma_{i,n}\in\mathcal C_{i,n}}
    F(\sigma_{1,n},\sigma_{2,n}) \leq\Bigl(\max_{i=1,2}\dim_{\mathbb R}(\mathcal D_{i,n})+1\Bigr)F_{\max}^n. 
\end{equation}
Moreover, we have for $i=1,2$, $\dim_{\mathbb R}(\mathcal D_{i,n})+1\leq\binom{n+d^2-1}{n}$, where $d$ is the dimension of the Hilbert space.
\end{lemma}
\begin{IEEEproof}
    The proof uses Carath\'eodory's theorem \cite[Theorem 0.0.1]{vershynin2025high} to write each element of $\mathcal C_{i,n}$ as a convex combination of $\dim_{\mathbb R}(\mathcal C_{i,n})+1$ elements of $\mathcal D_{i,n}$, then apply a version of \Cref{lemma:fidelity_for_finite} with $\dim_{\mathbb R}(\mathcal C_{i,n})=\dim_{\mathbb R}(\mathcal D_{i,n})$. For details, see Appendix \ref{sec:Proof of lemma:fidelity_for_infinite}.
\end{IEEEproof}

Comparing \Cref{lemma:fidelity_for_finite} and \Cref{lemma:fidelity_for_infinite}, we note that the latter yields a tighter bound whenever $\sqrt{\vert\mathcal D_1\vert\,\vert\mathcal D_2\vert} > \max_{i=1,2}\dim_{\mathbb R}(\mathcal D_{i,n})+1
$.
However, \Cref{lemma:fidelity_for_infinite} shows that the multiplicative factor $\dim_{\mathbb R}(\mathcal D_{i,n})+1$ scales as $\poly(n)$. Consequently, the following upper bound on the probability of error obtained by combining the above lemma with \eqref{eq:fidelity upper bound},
\begin{equation}\label{eq:ub_infinite_error}
 \hspace{-0.2cm}P_{e,\min}(p,\mathcal D_{1,n},\mathcal D_{2,n})\leq \sqrt{p(1-p)}\binom{n+d^2-1}{n}F_{\max}^n, 
\end{equation}
is not necessarily monotonically decreasing in $n$, and therefore does not capture the behaviour predicted by \Cref{lemma:error probability is non-increasing in n}. Hence, to derive analytical sample complexity in the infinite cardinality setting, we restrict the uncertainty sets to satisfy a max-fidelity constraint $F_{\max} \leq c$,  for some constant $c\in(0,1)$. This ensures that the polynomial prefactor admits a uniform upper bound over $n$, yielding exponential decay of the error probability. The following theorem then presents an upper bound on the sample complexity.
\begin{theorem}\label{th:infinite_qubit}
      Let $\mathcal D_{1}, \mathcal{D}_2$ be compact sets of quantum states,
possibly of infinite cardinality, satisfying $F_{\max}\leq c$ for some $c\in(0,1)$. For  $p,\delta \in (0,1)$, we have
\begin{align}
&n^\ast(\delta)\leq \biggl\lceil \frac{2\ln \Bigl( \frac{\sqrt{p(1-p)}K_{c,d}}{\delta}\Bigr)}{-\ln F_{\max}}\biggr\rceil,\label{eq:ub_infinite}
 \end{align}
where $K_{c,d}:= \binom{N+d^2-1}{N}c^{N/2}$ and $N:=\max\Bigl\{1,\Bigl\lceil\frac{d^2\sqrt{c}-1}{1-\sqrt{c}}\Bigr\rceil\Bigr\}$. Additionally, for fixed $p$, $c$ and $d$, \eqref{eq:symmetric_lb} and \eqref{eq:ub_infinite} together imply that the sample complexity scales as $ n^\ast(\delta)= \Theta \Bigl(\frac{\ln(1/\delta)}{-\ln F_{\max}}\Bigr)$.
\end{theorem}
\begin{IEEEproof}
Under the max-fidelity constraint $F_{\max} \leq c$, the relation in \eqref{eq:ub_infinite_error} can be further upper bounded as 
\begin{align*}
P_{e,\min}(p,\mathcal D_{1,n},\mathcal D_{2,n})
&\leq \sqrt{p(1-p)}\binom{n+d^2-1}{n}c^{n/2}F_{\max}^{n/2}\\
&\leq \sqrt{p(1-p)}K_{c,d}F_{\max}^{n/2},
\end{align*}
where $K_{c,d}$ and $N$ are defined as in the statement, and are independent of $n$. See Appendix \ref{Appendix:Extra analysis} for more details on this analysis. Choosing $n$ as in \eqref{eq:ub_infinite} ensures $P_{e,\min}(p,\mathcal D_{1,n},\mathcal D_{2,n})\leq \delta$, completing the proof.
\end{IEEEproof}

\section{Differentially Private Composite QHT}\label{sec:lqdp}
We now extend the previous results to the setting when the unknown quantum state is pre-processed by a noisy quantum channel before we receive it for testing. Specifically, we focus on the class  of $\varepsilon$-locally differentially private quantum (LDPQ) channels, defined as follows. 
\begin{definition}
 A CPTP map $\mathcal{M}$ is $\varepsilon$-LDPQ for $\varepsilon\geq0$ if
\begin{equation}
  \sup_{\rho, \sigma} E_{e^\varepsilon} \big(
 \mathcal{M} (\rho) \| \mathcal{M}(\sigma) \big) = 0,
\end{equation}
where $E_{\gamma}(\rho,\sigma)=\tr\left[(\rho-\gamma\sigma)_+\right]$ is the quantum hockey-stick divergence \cite{hirche2023quantum}.    
\end{definition}
Let $\mathcal M$ be an $\varepsilon$-LDPQ channel applied independently to each of the $n$-copies of the unknown quantum state. We then denote the locally private uncertainty sets as
\begin{equation*}
\mathcal D_{i,n}^{\mathcal M}
:= \{ \mathcal M(\rho_i)^{\otimes n} : \rho_i \in \mathcal D_i \},
\end{equation*}
for $i=1,2$, and define the $(\varepsilon,\delta)$-LDPQ sample complexity
\begin{align}
    n^\ast_\varepsilon(p,\mathcal D_{1}, \mathcal D_{2},\delta) := \!\inf_{\substack{\mathcal M\in \LDP_{\varepsilon}}} n^\ast(p,\mathcal D_{1}^{\mathcal M}, \mathcal D_2^{\mathcal M},\delta), 
\end{align}
where $\LDP_{\varepsilon}$ is the set of all $\varepsilon$-LDPQ channels. As before, we abbreviate the above notation as $n^*_{\varepsilon}(\delta)$, when the prior and the uncertainty sets are clear.

The following result provides a lower bound on this sample complexity, where we define $H_{1/2}(\rho\|\sigma) :=2\left(1-\tr[\sqrt{\rho}\sqrt{\sigma}]\right)$ as the Hellinger divergence of order $1/2$.
\begin{proposition}\label{prop:lb_sc_lqdp}
Let $\mathcal D_{1}, \mathcal{D}_2$ be compact sets of quantum states such that $H_{1/2}(\rho_1\|\rho_2) \leq 1$ for every $(\rho_1,\rho_2) \in \mathcal{D}_1 \times \mathcal D_2$. Then for $p,\delta \in (0,1)$ and $\varepsilon\geq 0$, we have 
\begin{align}
&\max \biggl \lbrace \frac{\bigl(1-\tfrac{\delta}{p}\bigr)^2}{e^{-\varepsilon}(e^\varepsilon-1)^2 \inf_{\rho_i\in\mathcal D_i, i=1,2} E_{1}(\rho_1\|\rho_2)^2}, \nonumber \\
&\frac{(e^{\varepsilon}+1)\ln\bigl(\frac{p(1-p)}{\delta}\bigr)}{2(e^{\varepsilon}-1)\inf_{\rho_i\in\mathcal D_i, i=1,2}H_{1/2}(\rho_1\|\rho_2)}\biggr \rbrace\leq  n^{\ast}_{\varepsilon}(\delta). \label{eq:lb_dp}
\end{align}
\end{proposition}
\begin{IEEEproof}
Using \cite[Lemma 1]{Cheng2024LDPQHT}, we have that $-\ln F(\rho_1,\rho_2) \leq-\ln\left(1-\frac{1}{2}H_{1/2}(\rho_1\|\rho_2)\right)\leq H_{1/2}(\rho_1\|\rho_2)$, where we used $-\ln(1-x)\leq2x$ for all $x\in[0,\frac{1}{2}]$. Applying this to the lower bound of \Cref{th:singleton}, and  using \cite[(4.12)]{Cheng2024LDPQHT} gives $$n^\ast_\varepsilon(p,\rho_1,\rho_2,\delta) \geq
\frac{(e^{\varepsilon}+1)\ln\bigl(\frac{p(1-p)}{\delta}\bigr)}{2(e^{\varepsilon}-1)H_{1/2}(\rho_1\|\rho_2)}.$$ Additionally, \cite[(4.25)]{Cheng2024LDPQHT} yields that $$n^\ast_\varepsilon(p,\rho_1,\rho_2,\delta) \geq \frac{(1-\delta/p)^2}{e^{-\varepsilon}(e^\varepsilon-1)^2E_1(\rho_1\|\rho_2)^2}.$$
\enlargethispage{-.07in}The proof then follows from noting that $ n^\ast_\varepsilon(p,\mathcal D_{1}, \mathcal D_{2},\delta) \ge n^\ast_\varepsilon(p,\rho_1,\rho_2,\delta)$ for any $(\rho_1,\rho_2) \in \mathcal{D}_1 \times \mathcal{D}_2$, whereby we get $ n^\ast_\varepsilon(p,\mathcal D_{1}, \mathcal D_{2},\delta) \ge \sup_{\rho_i\in\mathcal D_i} n^\ast_\varepsilon(p,\rho_1,\rho_2,\delta)$.
\end{IEEEproof}

The above result directly generalizes the sample complexity of locally differentially private simple QHT, recovering \cite[Theorems 4.6 and 4.9]{Cheng2024LDPQHT} for $p=0.5$, $\delta=0.1$ and $|\mathcal{D}_i|=1$ for $i=1,2$.

However, as in the previous section, the real challenge is to derive tight upper bounds on the sample complexity. To this end,
we derive the following result.
\begin{lemma}\label{lemma:ub_fidelity_lqdp}
Consider the setting of \Cref{prop:lb_sc_lqdp}. Then, the following inequality holds,
\begin{equation}
  \frac{1}{2}\biggl(\frac{e^\varepsilon-1}{e^\varepsilon+1}\biggr)^2\inf_{\rho_i\in\mathcal D_i} E_{1}(\rho_1\|\rho_2)^2 \leq -\inf_{\substack{\mathcal M\in \LDP_{\varepsilon}}}\ln F^{\mathcal{M}}_{\max}, 
\end{equation} where $F^{\mathcal{M}}_{\max}=\sup_{\rho_i\in\mathcal D_i, i=1,2}F(\mathcal M(\rho_1),\mathcal M(\rho_2))$.
\end{lemma}
\begin{IEEEproof}
    Let $\mathcal{B}$ be the  $\varepsilon$-LDP channel as constructed in the proof of \cite[Theorem 4.2]{Cheng2024LDPQHT}. Then, we have
\begin{align*}
-\inf_{\substack{\mathcal M\in \LDP_{\varepsilon}}}\ln F^{\mathcal{M}}_{\max} &\geq1-\inf_{\substack{\mathcal M\in \LDP_{\varepsilon}}}F^{\mathcal{M}}_{\max} \\ 
 &\geq  \frac{1}{2} \sup_{\substack{\mathcal M\in \LDP_{\varepsilon}}}\inf_{\rho_i\in\mathcal D_i} E_{1}(\mathcal M(\rho_1)\|\mathcal M(\rho_2))^2\\
 &\geq \frac{1}{2}\left(\frac{e^\varepsilon-1}{e^\varepsilon+1}\right)^2\inf_{\rho_i\in\mathcal D_i} E_{1}(\rho_1\|\rho_2)^2,
\end{align*}
where the first inequality uses  $1-x\leq -\ln x$ for all $x\in[0,1]$, the second  uses \cite[Equation 4.7]{Cheng2024LDPQHT}, and the last inequality  uses the specific channel $\mathcal{B}$ to evaluate $E_{1}(\mathcal B(\rho_1)\|\mathcal B(\rho_2))$.
\end{IEEEproof}
\Cref{lemma:ub_fidelity_lqdp} establishes a lower bound on $-\ln F_{\max}$ after $\varepsilon$-LDPQ pre-processing. Since all upper bounds derived in \Cref{sec:main} are monotonically decreasing functions in $-\ln F_{\max}$, combining \Cref{lemma:ub_fidelity_lqdp} with these expressions, yields valid upper bounds for the private setting. In particular, the following theorem elucidates this for finite cardinality sets.
\begin{theorem}
    Let $\mathcal{D}_1$, $\mathcal{D}_2$ be compact sets of quantum states satisfying $|\mathcal{D}_i|=m_i<\infty$ for $i=1,2$. For $\varepsilon \geq 0$, and $p, \delta \in(0,1)$, we have
    \begin{equation}
n^{\ast}_{\varepsilon}(\delta)  \leq
\left\lceil
\left(\frac{e^\varepsilon+1}{e^\varepsilon-1}\right)^2
\frac{2\ln\Bigl(\tfrac{\sqrt{m_1m_2p(1-p)}}{\delta}\Bigr)}{\inf_{\rho_i\in\mathcal D_i} E_{1}(\rho_1\|\rho_2)^2}
\right\rceil.    \label{eq:ub_dp}
 \end{equation}
Lastly, for fixed $p$ and $\delta$, \eqref{eq:lb_dp} and \eqref{eq:ub_dp} together imply that
\begin{multline}
\Omega \left( \frac{1}{e^{-\varepsilon}(e^\varepsilon-1)^2 \inf_{\rho_i\in\mathcal D_i} E_{1}(\rho_1\|\rho_2)^2}\right) \leq n^\ast_{\varepsilon}(\delta)\\\leq \mathcal{O}\left( \left(\frac{e^\varepsilon+1}{e^\varepsilon-1}\right)^2
\frac{\ln (m_1m_2)}{\inf_{\rho_i\in\mathcal D_i} E_{1}(\rho_1\|\rho_2)^2}\right).
\end{multline} 
\end{theorem}

\section{Conclusion}
\enlargethispage{-.07in}This work characterized the sample complexity of binary composite QHT, establishing bounds that highlight the roles of set cardinality, maximum pairwise fidelity, and Hilbert space dimensionality. We extended these results to the locally differentially private regime, providing a robust framework for state discrimination under privacy constraints. Future research may generalize this to $M$-ary composite hypotheses and composite channel discrimination, or refine the analysis through an extension to the asymmetric setting.

\bibliographystyle{IEEEtran}
\bibliography{references}
\appendix
\subsection{Proof of \Cref{trivial_cases}.}\label{Proof of  trivial cases}
We first show that, if $\rho_1\perp\rho_2$ (i.e. $\rho_1\rho_2=0$), then $\|p\rho_1-(1-p)\rho_2\|_1=1$. Note that $\rho_1,\rho_2$ are simultaneously diagonalisable, and have orthogonal supports. Thus $p\rho_1$ is the positive-part and $(1-p)\rho_2$ is the negative-part of the operator $(p\rho_1-(1-p)\rho_2)$. By a property of the trace-norm, we have
\begin{align*}
    \|p\rho_1-(1-p)\rho_2\|_1&=\tr(p\rho_1)+\tr((1-p)\rho_2)\\
    &=p\tr(\rho_1)+(1-p)\tr(\rho_2)\\
    &=p+(1-p)=1.
\end{align*}
Now assume that $\mathcal D_1\perp\mathcal D_2$, i.e. for all $\rho_1\in\mathcal D_1$ and $\rho_2\in\mathcal D_2$, $\rho_1\rho_2=0$. Since orthogonality of supports is preserved under convex combinations and tensor powers, this implies that for all $n\in\mathbb N$, $\mathcal C_{1,n}\perp\mathcal C_{2,n}$. Therefore, for all $\sigma_{i,n}\in\mathcal C_{i,n}$ we have
\begin{equation*}
    \|p\sigma_{1,n}-(1-p)\sigma_{2,n}\|_1=1.
\end{equation*}
Applying the above to \eqref{eq:prob_error_composite_explicit}, establishes that the error probability is $0$, thus a single sample suffices and $n^\ast(\delta)=1$. If $\delta\in[1/2,1]$, we can achieve this upper bound with a uniformly random guess, which carries an inherent error probability of $\frac12p+\frac12(1-p)=\frac12$. Thus, again $n^\ast(\delta)=1$. If $\delta\geq p^s(1-p)^{1-s}$ for some $s\in [0,1]$, note that using \eqref{eq:quantum_chernoff_bound}, we can upper bound the error probability by
\begin{align*}
    P_{e,\min}(p,\mathcal D_{1,n},\mathcal D_{2,n})&\leq\sup_{\sigma_{i,n}\in\mathcal C_{i,n}}p^s(1-p)^{1-s}\tr(\sigma_{1,n}^s\sigma_{2,n}^{1-s})\\
    &\leq p^s(1-p)^{1-s}.
\end{align*}
Thus, if $\delta\geq p^s(1-p)^{1-s}$, a single sample suffices, and $n^\ast(\delta)=1$. Finally, assume that $\mathcal D_1\cap\mathcal D_2\neq\emptyset$ and $\delta<\min\{p,1-p\}$. Then, there exists a state $\rho\in \mathcal D_1\cap\mathcal D_2$, such that both hypotheses correspond to the same state $\rho^{\otimes n}$. In this case, no measurement can distinguish the hypotheses. Therefore, the optimal strategy is to guess, yielding a minimum achievable error probability of $\min\{p,1-p\}$. However, $\delta<\min\{p,1-p\}$ so the error threshold can never be satisfied for any $n$, hence $n^\ast(\delta)=+\infty$.

\subsection{Proof of optimality of $s=0$ for $p\geq \frac{1}{2}$ in \Cref{prop:sample_complexity_qsvp}.}\label{sec:Proof of optimality in prop:sample_complexity_qsvp.}
Let
\begin{equation}
 f(s)=p^s(1-p)^{1-s}\tr((\proj\psi^{\otimes n})^s\sigma_{2,n}^{1-s}),
\end{equation}
The goal is to show that $f(s)$ is minimized at $s=0$ for $p\geq \frac{1}{2}$, that is
\begin{equation*}
\inf_{s\in[0,1]}f(s)=f(0).
\end{equation*}
Writing $\sigma_{2,n}=\sum_{i} \lambda\proj i$ i.e. in its spectral decomposition, and using that for all $s \in [0,1]$, $(\proj\psi^{\otimes n})^s=\proj\psi^{\otimes n}$ since $\proj\psi^{\otimes n}$ is a rank one projector, we have
\begin{align*}
 f(s)&=p^s(1-p)^{1-s}\bra\psi^{\otimes n}\sigma_{2,n}^{1-s}\ket\psi^{\otimes n}\\
 &=p^s(1-p)^{1-s}\sum_i\lambda_i^{1-s}\vert\braket{i}{\psi}^{\otimes n}\vert^2 \label{eq:qsvp_functon}.
\end{align*}
Define $Z(s):=\sum_i\lambda_i^{1-s}\vert\braket i\psi^{\otimes n}\vert^2$ as the Partition function, and the $s$-tilted distribution
\begin{equation}
\mu_s(i):=
\frac{\lambda_i^{1-s}\vert\braket i\psi^{\otimes n}\vert^2}
{Z(s)}
\end{equation}
Taking logarithms of $f(s)$, we have
\begin{equation*}
\ln f(s)=s\ln{p}+(1-s)\ln{(1-p)}+\ln Z(s)
\end{equation*}
and differentiating gives 
\begin{align*}
   \frac{d}{ds}\ln f(s)&= \ln\frac{p}{1-p}+\frac{Z'(s)}{Z(s)} \\ &= \ln\frac{p}{1-p} - \mathbb{E}_{\mu_s}[\ln\lambda].
\end{align*}
where the expectation is taken over the support of $\mu_s$, which only includes indices with $\lambda_i>0$. Since $0<\lambda_i\leq 1$, then $\ln (\lambda_i) \leq0$ and thus $- \mathbb{E}_{\mu_s}[\ln\lambda]\geq 0 $ with equality if and only if every $\lambda_i=1$ on the support of $\proj\psi^{\otimes n}$. This only occurs when $\sigma_{2,n}=\proj\psi^{\otimes n}$, however, $\mathcal D_1\cap\mathcal D_2=\emptyset$ so this is not possible. Therefore, $-\mathbb{E}_{\mu_s}[\ln\lambda]>0$. As well, since $p\geq\frac{1}{2}$, then $\ln\frac{p}{1-p}\geq0$. Altogether this gives $\frac{d}{ds}\ln f(s)>0$, hence $\ln f(s)$ is strictly increasing on $(0,1)$ and therefore so is $f(s)$. Lastly, $f(s)$ is continuous on $[0,1]$, since $Z(s)$ is a finite sum of continuous functions. Thus, we conclude that
\begin{align}
\inf_{s\in[0,1]} f(s)=f(0).
\end{align}
\subsection{Proof of \Cref{lemma:fidelity_for_infinite}.}\label{sec:Proof of lemma:fidelity_for_infinite}

The goal is to bound, $\dim_{\mathbb R}(\mathcal C_{i,n})$, i.e. the real (affine) dimensionality of $\mathcal C_{i,n}$, which allows us to apply Carath\'eodory's Theorem\cite[Theorem 0.0.1]{vershynin2025high} and then use \Cref{lemma:fidelity_for_finite}. Firstly, we define an important subspace of interest.
\begin{definition}
    Let $\mathcal B(\mathcal H)$ denote the set of all Hermitian operators, acting on Hilbert space $\mathcal H$. We define the subspace of permutation-invariant operators as
    \begin{equation}
        \mathcal S_n:=\{A\in\mathcal B((\mathbb C^d)^{\otimes n}):V(\pi)AV(\pi)^\dagger=A,\forall\pi\in\mathfrak S_n\},
    \end{equation}
    where $\mathfrak S_n$ is the symmetric group, i.e. the set of all permutations on $\{1,\ldots,n\}$, and $V(\pi)$ is the unitary matrix that permutes the $n$ copies of $(\mathbb C^d)^{\otimes n}$ according to $\pi$.
\end{definition}

Note that $\mathcal S_n$ is a real vector space (under addition and real scalar multiplication), which implies that $\dim_{\mathbb R}(\mathcal S_n)$ is the number of linearly independent vectors whose real
span equals that vector space. Since every operator of the form $\rho^{\otimes n}$ is permutation invariant, $\rho^{\otimes n}\in\mathcal S_n$ and hence $\mathcal C_{i,n} \subset \mathcal S_n$. Moreover, as all elements of $\mathcal C_{i,n}$ have unit trace, $\mathcal C_{i,n}$ lies in an affine hyperplane of codimension one in $\mathcal S_n$. Thus
\begin{equation}
   \dim_{\mathbb R}(\mathcal C_{i,n})\leq\dim_{\mathbb R}(\mathcal S_n)-1.
\end{equation}
Next, to find $\dim_{\mathbb R}(\mathcal S_n)$, we can follow the argument of \cite[Theorem 17]{mele2024introduction}, which states that the dimensionality of the symmetric subspace of permutation-invariant vectors in $(\mathbb C^d)^{\otimes n}$ is $\binom{n+d-1}{d-1}$ (assuming that complex linear combinations are allowed). Since we work with density operators, the relevant single copy space, is the space of $d\times d$ Hermitian matrices, which can easily be verified to be a real vector space of dimension $d^2$. Therefore, changing $d$ to $d^2$, and considering only real linear combinations (which is possible since we have a real vector space), we obtain
\begin{equation}
    \dim_{\mathbb R}(\mathcal S_n)=\binom{n+d^2-1}{d^2-1},
\end{equation}
and altogether we get



\begin{equation}
    \dim_{\mathbb R}(\mathcal C_{i,n})\leq\binom{n+d^2-1}{d^2-1}-1.
\end{equation}
Finally, to obtain \eqref{eq:fidelity_for_infinite}, applying Carath\'eodory's theorem, we have that any $\sigma_{i,n}\in\mathcal{C}_{i,n}$ can be written as a convex combination of at most $\dim_{\mathbb R}(\mathcal C_{i,n})+1$ elements in $\mathcal D_{i,n}$. Hence, applying \Cref{lemma:fidelity_for_finite} for $m_i=\dim_{\mathbb R}(\mathcal C_{i,n})+1$ and using that $\dim_{\mathbb R}(\mathcal C_{i,n})=\dim_{\mathbb R}(\mathcal D_{i,n})$, yields
\begin{equation}
\hspace{-0.215cm}\sup_{\sigma_{i,n}\in\mathcal C_{i,n}}
    F(\sigma_{1,n},\sigma_{2,n}) \leq\left(\max_{i=1,2}\dim_{\mathbb R}(\mathcal D_{i,n})+1\right)F_{\max}^n. 
\end{equation}
\subsection{Extra Analysis of \Cref{th:infinite_qubit}}\label{Appendix:Extra analysis}
For fixed $x>0$, define 
\begin{equation*}
    g(n):=\binom{n+d^2-1}{n}e^{-xn}\text{ and }  N:= \min \left\{n\in \mathbb N:R(n)\leq 1\right\}
\end{equation*}
where
\[
 R(n):= \frac{g(n+1)}{g(n)}.
\] 
We will now show that $g(n)$ is unimodal, that is, it monotonically increases for all $n<N$, and then  monotonically decreases for all $n\geq N$. Furthermore , it attains this maximum at
\begin{equation}
N=\max\left\{1, \left\lceil\frac{d^2-e^x}{e^x-1}\right\rceil \right\}.
\end{equation}
Note that 
\begin{align*}
   R(n)&:=\frac{g(n+1)}{g(n)}
   =\frac{ \binom{(n+1)+d^2-1}{n+1}e^{-x(n+1)}}{\binom{n+d^2-1}{n}e^{-xn}}= \frac{n+d^2}{n+1}e^{-x},
\end{align*}
and hence $R(n)$ is strictly decreasing for all $n$. Thus there exists at most one unique turning point. Moreover, since 
\[
\lim_{n\to \infty}R(n)=e^{-x}<1,
\]
the set $\left\{n\in \mathbb N:R(n)\leq 1\right\}$ is non-empty and hence $N$ is well defined. Therefore, by minimality of $N$, we have 
\[
R(n)>1 \text{ for all } n < N \text{ and } R(n)\leq 1 \text{ for all } n \geq N.
\]
It follows that 
\[
R(n)>1 \implies g(n+1)> g(n) 
\]
and 
\[
R(n)\leq1 \implies g(n+1)\leq g(n).
\]
Hence $g(n)$ monotonically increases for all $n<N$, and monotonically decreases for all $n\geq N$. Furthermore, taking $R(N)\leq 1$ (i.e., the point at which $g(n)$ becomes monotonically decreasing), and rearranging for $N$ gives
\begin{align*}
    N\geq\frac{d^2-e^x}{e^x-1}. \implies N=\max\left\{1, \left\lceil\frac{d^2-e^x}{e^x-1}\right\rceil \right\}.
\end{align*}
Choosing $x=-\frac{1}{2}\ln c$ gives  $K_{c,d}:=\sup_{n \le N} \binom{n+d^2-1}{n}c^{n/2}$ and $N:=\max\{1,\lceil\frac{d^2\sqrt{c}-1}{1-\sqrt{c}}\rceil\}$.
\end{document}